\documentclass[12pt]{article}

\usepackage{times}
\usepackage[numbers]{natbib}
\usepackage{hyperref}
\usepackage{subfigure}
\usepackage{amsmath,amsfonts,amssymb,array}
\usepackage{mathabx}
\usepackage{mathtools}
\usepackage[utf8]{inputenc}
\usepackage{graphicx}
\usepackage{enumerate}
\usepackage{multirow}
\usepackage{xcolor}
\usepackage{booktabs}

\newenvironment{sciparagraph}{%
\begin{quote} \bf}
{\end{quote}}

\newcounter{lastnote}

\usepackage[labelfont=bf]{caption}
\title{Basin stability and limit cycles in a conceptual model for climate tipping cascades}

\author
{Nico Wunderling,$^{1,2,3\footnote{Correspondences should be addressed to nico.wunderling@pik-potsdam.de or jonathan.donges@pik-potsdam.de}}$  Maximilian Gelbrecht,$^{3,4}$ Ricarda Winkelmann$^{1,2}$,\\ Jürgen Kurths$^{3,4,5}$ \& Jonathan F. Donges$^{1,6 \footnotemark[1]}$\\
\\
\normalsize{$^{1}$Earth System Analysis, Potsdam Institute for Climate Impact Research (PIK),}\\
\normalsize{ Member of the Leibniz Association, 14473 Potsdam, Germany}\\
\normalsize{$^{2}$Institute of Physics and Astronomy, University of Potsdam, 14476 Potsdam, Germany}\\
\normalsize{$^{3}$Department of Physics, Humboldt University of Berlin, 12489 Berlin, Germany}\\
\normalsize{$^{4}$Complexity Science, Potsdam Institute for Climate Impact Research (PIK),}\\
\normalsize{ Member of the Leibniz Association, 14473 Potsdam, Germany}\\
\normalsize{$^{5}$Lobachevsky State University Nizhny Novgorod, Nizhny Novgorod, Russia,}\\
\normalsize{$^{6}$Stockholm Resilience Centre, Stockholm University, Stockholm, SE-10691, Sweden}\\
}

\date{}

\begin{document} 

% Double-space the manuscript.

\baselineskip24pt

% Make the title.

\maketitle 
% Place your abstract within the special {sciabstract} environment.
\newpage

%------------------------------------------
\begin{sciparagraph}
Tipping elements in the climate system are large-scale subregions of the Earth that might possess threshold behavior under global warming with large potential impacts on human societies. Here, we study a subset of five tipping elements and their interactions in a conceptual and easily extendable framework: the Greenland and West Antarctic Ice Sheets, the Atlantic Meridional Overturning Circulation (AMOC), the El-Ni{\~n}o Southern Oscillation (ENSO) and the Amazon rainforest. In this nonlinear and multistable system, we perform a basin stability analysis to detect its stable states and their associated Earth system resilience. \textcolor{black}{By combining these two methodologies with a large-scale Monte Carlo approach, we are able to propagate the many uncertainties associated with the critical temperature thresholds and the interaction strengths of the tipping elements. Using this approach, we perform a system-wide and comprehensive robustness analysis with more than 3.5~billion ensemble members.} Further, we investigate dynamic regimes where some of the states lose stability and oscillations appear using a newly developed basin bifurcation analysis methodology. \textcolor{black}{Our results reveal that the state of four or five tipped elements has the largest basin volume for large levels of global warming beyond 4~$^\circ$C above pre-industrial climate conditions, representing a highly undesired state where a majority of the tipping elements reside in the transitioned regime.} For lower levels of warming, states including disintegrated ice sheets on West Antarctica and Greenland have higher basin volume than other state configurations. \textcolor{black}{Therefore in our model, we find that the large ice sheets are of particular importance for Earth system resilience.} We also detect the emergence of limit cycles for 0.6\% of all ensemble members at rare parameter combinations. Such limit cycle oscillations mainly occur between the Greenland Ice Sheet and AMOC (86\%), due to their negative feedback coupling. These limit cycles point to possibly dangerous internal modes of variability in the climate system that could have played a role in paleoclimatic dynamics such as those unfolding during the Pleistocene ice age cycles.
\end{sciparagraph}\newpage\clearpage

\section{Introduction}
During the last decades, the field of tipping elements has become a major point of interest in complex systems and network science~\citep{scheffer2009critical,watts2002simple}. They have been used in the description of various fields such as in financial markets, technological progress, ecology or in climate science~\citep[e.g.,][]{may2008ecology,herbig1991cusp,scheffer2001catastrophic,lenton2008tipping}. Tipping elements can interact across scales in space and time~\citep{rocha2018cascading} which could potentially lead to catastrophic domino effects~\citep{brummitt2015coupled} or, for instance, lead to a hothouse cliamte state in the case of climate tipping elements~\citep{steffen2018trajectories}. 

In the climate system, tipping elements are subregions of the Earth system that can exhibit threshold behavior, where a small forcing perturbation can be sufficient to invoke a strong non-linear response of the system that can qualitatively change the state of the whole region or system due to internal, self-enforcing feedbacks~\citep{lenton2008tipping}. Climate tipping elements comprise systems from the cryosphere (e.g. Greenland, Antarctic Ice Sheet, Permafrost), the biosphere (e.g. Amazon rainforest, coral reefs) and large-scale circulation systems (e.g. Monsoon systems, Atlantic Meridional Overturning Circulation)~\citep{lenton2008tipping,schellnhuber2016right}. Their potential tipping to alternative states would be associated with severe impacts on the biosphere and threaten human societies~\citep{lenton2019climate}.\\

It has been suggested that several climate tipping elements are at risk or on the way of transgressing into an undesired state even at global warming levels below the 2.0$^\circ$C goal of the Paris Agreement~\citep{schellnhuber2016right,lenton2019climate,wang2020esd}. Among others, tipping elements that already show warning signals of degradation at present times~\citep{lenton2019climate,wang2020esd} are: the West Antarctic Ice Sheet where parts in the Amundsen Bay (Pine Island \& Thwaites region) are suspected to have been destabilized~\citep{joughin2014marine,favier2014retreat,rosier2020tipping}, the AMOC which experienced a major slowdown of 15\% from 1950 to now~\citep{caesar2018observed}, the Amazon rainforest which might approach a tipping point due to climate change and deforestation~\citep{lovejoy2019amazon}. Critical deforestation ratios might lie between 20 to 40\%, where current deforestation is reaching 20\%~\citep{lovejoy2019amazon,nobre2016land}. Furthermore, the Greenland Ice Sheet loses mass at an accelerating pace~\citep{stocker2013climate,zwally2011greenland} and the frequency of major El-Ni\~{n}o events are suggested to increase twofold and strong ENSO effects will occur more often as global warming continues~\citep{cai2014increasing,wang2019historical}. However, others highlight that large uncertainties are related to future changes of ENSO and whether major El-Ni\~{n}o events will become more frequent or intense under global warming~\citep{kim2014response,collins2010impact}. %\textcolor{red}{COMMENT: Here, we could also cite Niklas Boers' work on Early warning signals Greenland and AMOC (unfortunately not published yet to my knowledge).}

Furthermore, contradicting a common misunderstanding, tipping elements do not necessarily tip immediately after the crossing of their tipping point, but their tipping time trajectory might take very long and appear smooth~\citep{hughes2013multiscale}. For instance for the large ice sheets, the disintegration time scale could be on the order of several centuries up to millennia as has been suggested by modeling studies ~\citep{levermann2016simple,winkelmann2015combustion,robinson2012multistability}.

For most of the tipping elements, there is a critical temperature range at which they are suspected to leave their current safe state separating the climate tipping elements into three groups~\citep{schellnhuber2016right}. The first group comprises elements that might transgress their state within the limits of the Paris Agreement (Paris, 2015) of 2~$^\circ$C above pre-industrial and with that, these are the most vulnerable climate tipping elements with respect to global warming. This group contains mainly cryosphere elements (Arctic summer sea ice, West Antarctic Ice Sheet, Greenland Ice Sheet and Alpine glaciers) as well as the Coral reefs that are likely to be lost even when global warming is restricted to 2~$^\circ$C above pre-industrial. The second group might tip at temperatures above 3~$^\circ$C (for instance the Amazon rainforest, AMOC or ENSO) and the most resilient group only at temperatures around 5~$^\circ$C above pre-industrial or higher (e.g., parts of the Antarctic ice sheet, permafrost or Arctic winter sea ice).\\
 
However, the tipping elements in the climate system are not independent of each other, but connected~\citep{kriegler2009imprecise,lenton2019climate} and the knowledge about the exact interaction structure is sparse and partially based on experts that, for instance, suggested an interaction structure, including sign and strength, for a subset of five tipping elements: The Greenland Ice Sheet, the West Antarctic Ice Sheet, the AMOC, the Amazon rainforest and the ENSO~\citep{kriegler2009imprecise}. Behind each connection between two tipping elements within this subset, there is a physical process or set of processes (see Tab.~\ref{tab:appendix_1}). For instance the impact of the Greenland Ice Sheet on the AMOC due to freshwater input from melting ice slows down the AMOC on the one hand and a weakening AMOC on the other hand cools latitudes in the northern hemisphere. Note that this subset network of tipping is neither complete in the number and selection of tipping elements, nor is it comprehensive in the possible connection pathways and their potential strength between the tipping elements. There are also earlier investigations on tipping points~\citep{cai2015environmental} and the interaction of tipping points~\citep{cai2016risk,lemoine2016economics} in the context of economic damage and the social cost of carbon using further developed versions of the integrated assessment model DICE~\citep{nordhaus2014question,nordhaus2014estimates}. Here, Cai et al. (2016)~\citep{cai2016risk} explicitly base their findings on the interactions of tipping elements from the expert elicitation in Kriegler et al. (2009)~\citep{kriegler2009imprecise}.\\

Following the elaborations above, in this work we aim at investigating the resilience of various attractors for interacting climate tipping elements and we want to elucidate the role that different tipping cascades have in that regard. The approach put forward here can easily be adapted to more tipping elements and further interaction structures once they are more comprehensively understood~\citep[see also][]{wunderling2020interacting}.

\begin{figure}[htbp]
\centering
\includegraphics[width=0.8\textwidth]{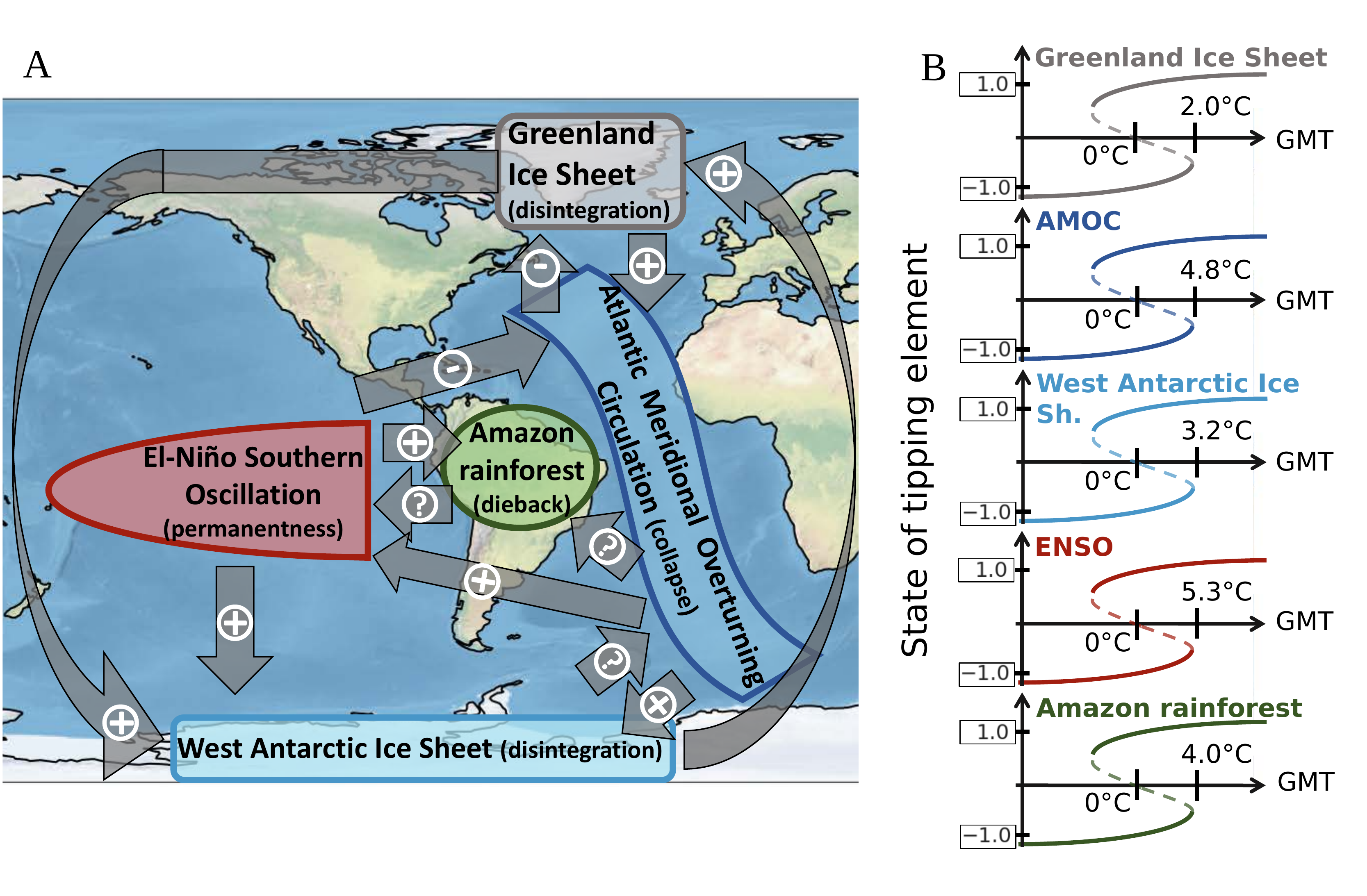}
\caption{A) World map with connections shown for five tipping elements where the interaction structure is known from an expert elicitation~\citep{kriegler2009imprecise,lenton2013origin}. Each link represents a physical mechanism and has a certain strength (see supplementary Tab.~\ref{tab:appendix_1}). A positive arrow implies an effect that drives the tipping element closer to its tipping point, a negative arrow drives the tipping element away from its tipping point and a question mark stands for an unclear direction. B) (Fold) Bifurcation diagram of each of the tipping elements without coupling. On the x-axis, the average global warming required for tipping is shown in degrees above pre-industrial global mean temperature levels for the respective tipping element. }
\label{fig:one}
\end{figure}

We explore the stability landscape and the dynamics of a subset of five tipping elements represented by normal form fold bifurcations based on known interactions across scales in time and space between these tipping elements (Fig.~\ref{fig:one})~\citep{kriegler2009imprecise}. These five tipping elements are the Greenland and West Antarctic Ice Sheets, the AMOC, the ENSO and the Amazon rainforest~\citep{kriegler2009imprecise}. We introduce the model of interacting tipping elements in section. We use the concept of basin stability~\citep{menck2013basin} in order to determine the basin sizes of various attractors of this multistable system (see section~\ref{methods}). Based on a large-scale Monte Carlo ensemble, this methodology gives an estimate how stable and resilient various attractors are. It has been applied to many dynamic systems before such as power grids, neuronal models and further nonlinear systems~\citep{schultz2014detours,leng2016basin,mitra2017multiple,schultz2017potentials,hellmann2016survivability}. Furthermore, especially for larger coupling strengths, Hopf bifurcations can occur, thus invoking oscillatory limit-cycle solutions of the model. For the detection and quantification of these types of limit cycle attractors, we apply a newly developed bifurcation algorithm that is able to identify different dynamical properties in complex systems: the Monte Carlo Basin Bifurcation analysis (MCBB)~\citep{gelbrecht2020monte}.

In the following, we first introduce the methodological approach of this work: the model of interacting tipping elements (section~\ref{met:model}), the basin stability approach (section~\ref{met:basin}), the construction of the large-scale Monte Carlo ensemble (section~\ref{met:monte}) and the Monte Carlo Basin Bifurcation analysis (section~\ref{met:mcbb}). Then, we evaluate the basin volume in our model (section~\ref{res:basin}) and quantify the occurrence of limit cycles in our model (section~\ref{res:osci}). Lastly, the results with respect to the climate system are discussed and summarized in sections~\ref{disc} and~\ref{conc}.\\

\section{Methods}
\label{methods}
For the purpose of investigating the dynamical properties of the subset of five tipping elements, we further developed a conceptual network approach that is fully dynamic and captures the main nonlinear dynamical properties of tipping elements~\citep{wunderling2020interacting,wunderling2020motifs,kronke2020dynamics}. The actual physical processes behind the tipping elements are not explicitly modeled to maintain an accessible and controllable structure. The modeling of complex systems using conceptual approaches is a popular tool and has been successfully applied to, among others, ecology, social systems or epidemiology~\citep{scheffer2001catastrophic,brockmann2013hidden,wiedermann2020network}.

\subsection{Tipping elements and interactions}
\label{met:interactions}
\textcolor{black}{Given that, despite major advances, current EMICs (Earth system models of intermediate complexity) and GCMs (global circulation models) are not yet able to fully represent the nonlinear behavior of some Earth system components together with their interactions, but physics based models and equations as well as paleo climate observations suggest the existence of such properties for many tipping elements as for instance the Greenland and (West) Antarctic Ice Sheet~\citep{levermann2016simple,robinson2012multistability,ridley2010thresholds,garbe2020hysteresis}, the AMOC~\citep{stommel1961thermohaline,rahmstorf2005thermohaline,hawkins2011bistability,mecking2016stable,wood2019observable}, the Amazon rainforest~\citep{oyama2003new,cox2004amazonian,hirota2011global,van2014tipping,staal2015synergistic} or the ENSO~\citep{timmermann2005enso,dekker2018cascading}, the conceptual approach chosen here demonstrates an option how to model interactions between tipping elements.} Thus, we put this model forward as a first step towards a more process-detailed assessment of tipping elements and their interactions. This also emphasizes that future research could focus on developing more complex, emulator- or EMIC/GCM-like models of tipping elements to investigate their nonlinear interplay such as has recently been developed for the Antarctic Ice Sheet~\citep{garbe2020hysteresis}. \\

\textcolor{black}{While we have described why we use a conceptual description of the main dynamics of the tipping elements directly above, we outline the physical mechanisms of there interactions hereafter.} \textcolor{black}{Although some interactions between the tipping elements are better understood and evaluated than others, we describe the physical mechanisms of all of them in the following separated into destabilizing ($+$), stabilizing ($-$) and unclear links (?) (see Fig.~\ref{fig:one}). This description aims to provide a basic physical understanding, but cannot resolve the problem of how strong exactly each of these interactions is. Therefore, the interaction structure is kept as described later in Eq.~\ref{eq:one} with the multiplicative interaction strength factor $d$.}
\textcolor{black}{
\begin{enumerate}
    \item Destabilizing interactions: \\
    I) Greenland Ice Sheet $\rightarrow$ AMOC: When the Greenland Ice Sheet starts to melt, it has a diminishing influence on the overturning strength of the AMOC due to freshwater input into the North Atlantic. This has been observed in modeling studies~\citep{caesar2018observed,rahmstorf2005thermohaline,jungclaus2006will,driesschaert2007modeling} and in observations~\citep{robson2014atlantic}.\\
    II) AMOC $\rightarrow$ West Antarctic Ice Sheet: When the AMOC collapses, sea surface temperature anomalies arise due to the collapse of the northward heat transport of the AMOC. This results in a cold north and a warm south of the equator as shown by modeling studies~\citep{stouffer2006investigating,timmermann2007influence,vellinga2002global,weijer2019stability}.\\
    III) Greenland Ice Sheet $\rightarrow$ West Antarctic Ice Sheet (and vice versa): The shift of grounding lines due to changing sea level is a well-known phenomenon from tidal changes~\citep[e.g.][]{sayag2013elastic}. Thus, if the sea level rises due to global warming, the floating ice shelves could be lifted which is likely to result in grounding line retreat. Furthermore, gravitational changes as well as elastic and rotational effects might then amplify the sea level change if one of the large ice sheets disintegrates first because the gravitational attraction then only emanates from the other, remaining ice sheet~\citep{mitrovica2009sea,kopp2010impact}. The effect would be stronger if Greenland melts first since the West Antarctic Ice Sheet has more marine terminating glaciers and ice shelves.\\
    IV) AMOC $\rightarrow$ ENSO: There are two opposing effects that have proposed, which describe how the AMOC might influence the ENSO: (i) It has been suggested that oceanic Kelvin waves originate from a colder North Atlantic and travel southward. Then, in western Africa, Rossby waves would be emitted towards the north and the south, which are then translated back into Kelvin waves that travel into the Pacific ocean. This effect would deepen the Pacific thermocline and weaken the amplitude of ENSO~\citep{timmermann2005enso}. (ii) With a weaker AMOC, the northern tropical Atlantic would in turn become cooler and northerly trade winds would be intensified over the northeastern tropical Pacific. It has been argued that this could result in a southward displacement of the Pacific ITCZ leading to a sea surface temperature anomaly~\citep{zhang2005simulated}. At the same time, it is argued that Rossby waves are sent into the northeast tropical Pacific~\citep{dong2005mechanism}. This would intesify ENSO due to wind stress interaction from AMOC. Overall, it is believed that mechanism (ii) is stronger than mechanism (i)~\citep{timmermann2005enso}. Furthermore, it can be extracted from complex Earth system models that a decrease in AMOC intensity indeed strengthens the variability of ENSO~\citep{dekker2018cascading,sterl2008can}.\\
    V) ENSO $\rightarrow$ Amazon rainforest: Literature studies suggest that droughts related to climate variabilities such as the El-Ni\~{n}o Southern Oscillation can affect the stability of the Amazon rainforest~\citep{holmgren2013effects,holmgren2006extreme,malhi2004spatial}. Using an EMIC, it has been found that a permanent El-Ni\~{n}o state would endanger substantial portions of the Amazon basin due to a reorganization and reduction of water access in the South American tropics via teleconnections~\citep{duque2019tipping}.\\
    VI) ENSO $\rightarrow$ West Antarctic Ice Sheet: The interaction between ENSO and the West Antarctic Ice Sheet is one of the least certain interactions as has already been stated in Kriegler et al (2009)~\citep{kriegler2009imprecise}. Nevertheless, there are hints for warming oceanic effects from El-Ni\~{n}o in the Amundsen and Ross Sea region, while La Ni\~{n}a would cool this region. At the same time, atmospheric effects could have an opposite effect, which would offset the oceanic effect~\citep{bertler2006opposing}. Besides that, it has been found with satellite observations that ice shelves gain height, but yet lose mass during El-Ni\~{n}o events in the Amundsen and Ross Sea region~\citep{paolo2018response}. While the primary driver of melt in West Antarctica is the warm ocean water below ice shelves, an extended period of surface melting has been observed during January 2016, which is likely promoted by the strong El-Ni\~{n}o event in this year~\citep{nicolas2017january}. Since it is expected that the frequency of major El-Ni\~{n}o events will increase during climate change~\citep{cai2014increasing}, we set this interaction positive (see Tab.~\ref{tab:appendix_1} and Fig.~\ref{fig:one}).\\
    \item Stabilizing interactions:\\
    I) AMOC $\rightarrow$ Greenland Ice Sheet: For a decreasing overturning strength of the AMOC, the northern hemisphere is cooled since the heat transport towards the North Atlantic would be weakened. This has been observed in modeling studies~\citep{caesar2018observed,stouffer2006investigating,timmermann2007influence,jackson2015global}.\\
    II) ENSO $\rightarrow$ AMOC: Using reanalysis data, evidence has been found that the transport of water vapor out of the tropical Atlantic is enhanced~\citep{schmittner2000enhanced}. Comparing La Ni\~{n}a and El-Ni\~{n}o conditions, it was found in this study that El-Ni\~{n}o conditions lead to a stronger northern AMOC on a multi-decadal timescale. However, another study questions this finding and does not find a strong impact on the deepwater formation from AMOC~\citep{spence2006impact}. Therefore, this interaction is less well established from literature and therefore considered of low strength, but with a negative sign (see Fig.~\ref{fig:one} and Tab.~\ref{tab:appendix_1}).\\
    \item Unclear interaction direction:\\
    I) AMOC $\rightarrow$ Amazon rainforest: When the AMOC shuts down, the intertropical convergence zone (ITCZ) is likely dislocated southward, leading to large changes in seasonal precipitation on a local to regional degree. This might then impact parts of the Amazon rainforest~\citep{weijer2019stability,jackson2015global,parsons2014influence}. Still, it is unknown as to whether this interaction is positive or negative and might differ from region to region. Therefore, this link is set as unclear (see Fig.~\ref{fig:one}).\\
    II) West Antarctic Ice Sheet $\rightarrow$ AMOC: A literature study using a coupled ocean-atmosphere model found a decrease in the AMOC for high freshwater inputs from the West Antarctic Ice Sheet~\citep{seidov2005there}. However, another study detected a stabilization of the AMOC if influenced by freshwater input from West Antarctica. This is ascribed to the effects from the bipolar ocean seesaw due to decreasing Antarctic Bottom Water formation~\citep{swingedouw2008antarctic}. With an EMIC, is has been found from using freshwater input experiments into the Southern Ocean that different processes could enhance or slow down the AMOC~\citep{swingedouw2009impact}: (i) The deep water adjustments via the bipolar ocean seesaw tend to intensify the NADW formation. (ii) The NADW is strengthened by southern hemispheric wind increase representing an ocean-atmosphere interaction. (iii) Salinity anomalies from the Southern Ocean are distributed to the North Atlantic weakening the NADW~\citep[compare to][]{seidov2005there}. Overall, the processes (i) and (iii) strengthen the AMOC and process (ii) weakens it. However, the exact time scale and efficiencies of these processes have been rated unknown as of yet~\citep{swingedouw2009impact}.\\
    III) Amazon rainforest $\rightarrow$ ENSO: Under a dieback of the Amazon rainforest, the moisture supply to the atmosphere will significantly change, also since the atmospheric moisture recycling feedback over the Amazon basin would break down~\citep{aragao2012environmental,boers2017deforestation,zemp2017self}. However, it is unclear whether and to which extent this would then impact ENSO.
\end{enumerate}
}

\subsection{Model}
\label{met:model}
In our conceptual model, we divide the dynamics $x_i$ of the considered tipping elements $i$ into their individual dynamics $f_i\left(x_i\right)$ and a direct interaction term $g_i(\vec{x}) \equiv g_i(x)$. This yields
\begin{equation}
    \tau_i \dot{x}_i = f_i\left(x_i\right) + g_i(x),
\label{eq:basic}
\end{equation}
where $\tau_i$ is the typical time that passes when a tipping element undergoes a critical transition from one state to another. We model the individual dynamics of each of the tipping elements with the general tipping approach (CUSP equation~\citep{brummitt2015coupled,abraham1991computational})
\begin{equation}
    f_i\left(x_i\right) = -a_i x_i^3 + b_i x_i + c_i \hspace{2cm} a_i, b_i, c_i \in \mathbb{R},
\label{eq:cusp}
\end{equation}
where $a_i>0$ and $b_i>0$. Assuming additive separability of the interactions between the tipping elements and linear interactions, the interaction term $g_i(x)$ becomes
\begin{equation}
    g_i(x) = \sum_j g_{ij}(x_i, x_j) \mathrel{\mathop{=}^{\mathrm{linear}}_{\mathrm{interactions}}} \sum_{j} A_{ij} x_j.
\end{equation}
Here, $A_{ij}$ is the interaction structure and strength, which is set to zero if there is no connection between the tipping elements $i$ and $j$. Altogether, Eq.~\ref{eq:basic} becomes
\begin{equation}
    \tau_i \dot{x_i} = -a_i x_i^3 + b_i x_i + c_i + \sum_{j} A_{ij} x_j.
\end{equation}
Each tipping element $x_i$ following this equation possesses two fold bifurcations at $\pm \sqrt{(4 a_i^3)/(27 b_i)}$ and has already been investigated in theoretical works on tipping cascades~\citep{abraham1991computational}, but also in various contexts where nonlinear behavior is important as for instance in policy, environmental issues, economy or climate~\citep{brummitt2015coupled,klose2020emergence}. For these equations exist a framework that allows to investigate tipping cascades on larger networks with regard to their interaction structure in the network as well as microstructures that are decisive for finding emergent tipping cascades~\citep{wunderling2020motifs,kronke2020dynamics}. \\

\textcolor{black}{In our model, we specify the interaction structure and strength term $A_{ij}$ by setting it equal to a multiplicative factor $d$ times the actual link strength $s_{ij}$ between each pair of tipping elements. Therefore, $A_{ij} = d/10 \cdot s_{ij}$. The link strength values $s_{ij}$ are taken from the expert elicitation~\citep{kriegler2009imprecise}. The factor $1/10$ is used for normalization reasons since then $d \in [0, 1]$. If we now additionally set $a=1, b=1$ and $c_i = \left( \sqrt{4/27}/T_\text{limit,\ i} \right) \cdot \Delta \text{GMT}$, the tipping elements are described by the following nonlinear, ordinary differential equation (all parameters of Eq.~\ref{eq:one} are explained in the Tabs.~\ref{tab:appendix_1} and~\ref{tab:appendix_2})}
\begin{equation}
\frac{dx_i}{dt} = \left[ - x_i^3 + x_i + \frac{\sqrt{4/27}}{T_\text{limit,\ i}} \cdot \Delta \text{GMT} + \frac{d}{10} \cdot \sum_{\substack{j\\j \neq i}} s_{ij} \left( x_j + 1 \right) \right] \frac{1}{\tau_i}.
\label{eq:one}
\end{equation}
Here, $x_i$ is the state of the tipping element (see Fig.~\ref{fig:one}B) and $i$ stands for the considered tipping elements $i={\text{Greenland\ Ice\ Sheet}, \text{West\ Antarctic\ Ice\ Sheet}, \text{AMOC}, \text{ENSO}, \text{Amazon\ rainforest}}$. We choose these five tipping elements since their interaction structure is known from an expert elicitation~\citep{kriegler2009imprecise}. The increase of the global mean temperature above pre-industrial is denoted by $\Delta \text{GMT}$, $T_\text{limit,\ i}$ is the critical temperature threshold of the respective element. The last term is the coupling term, where $d$ is a general multiplicator that determines the strength of the interaction term in comparison to the other, individual dynamics terms. The parameter d is varied between 0, meaning no interactions, and 1, where the interactions become as important as the individual dynamics. Following this, one might tend to assume that the individual dynamics of the tipping element influences the tipping element more than the interaction effect. This might make smaller coupling parameters more realistic than higher ones. In Eq.~\ref{eq:one}, $s_{ij}$ is the link strength that is based on the expert elicitation~\citep{kriegler2009imprecise} and $\tau_i$ is a typical timescale at which a certain tipping element transgresses its state.\\

This typical tipping time scale ranges from decades for the Amazon rainforest to several millennia for the large ice sheets (see Appendix Tab.~\ref{tab:appendix_2}). Then, our system of differential equations is integrated forward in time using \textit{scipy.odeint}~\citep{virtanen2020scipy} until more than 20 times the Greenland Ice Sheet's typical transition time scale has passed. This is equal to 100,000 years simulation time. This is the time when equilibrium is reached in the simulations. However, we are not intending to compute an exact time scale for tipping or tipping cascades here, but we are rather interested in the system's attractors and their stability properties. This is why we denote model years in arbitrary units instead of giving an exact time, also since this would be beyond the scope of this conceptual model (see Fig.~\ref{fig:one}A). Note that we adapted the link from ENSO to AMOC from uncertain to negative compared to the original results of the expert elicitation on tipping element interactions~\citep{kriegler2009imprecise} since there is only a dampening process known in literature~\citep{lenton2013origin}. 

There are considerable uncertainties associated with this approach, especially with the critical temperature at which a certain tipping element transgresses its state $T_\text{limit,\ i}$ as well as in the strength of the interactions $s_{ij}$. The uncertainties of these two parameters are shown in the Appendix Tabs.~\ref{tab:appendix_1} and~\ref{tab:appendix_2}. Thus, with Eq.~\ref{eq:one}, we model tipping events and cascades under certain conditions of global warming (GMT) and the interaction strength ($d$).

\subsection{Basin stability}
\label{met:basin}
We are interested in the stability properties of different attractors within the state space. An appropriate tool to investigate the stability landscape of such states is the so-called \textit{basin stability}~\citep{menck2013basin,mitra2017multiple}. Basin stability is a nonlinear stability measure for the resilience of an attractor to disturbances. \textcolor{black}{Where traditional measures such as the computation of Lyapunov exponents or Master stability functions rely on linear approximations in reaction to small perturbations~\citep{nishikawa2006synchronization,pecora1998master}, basin stability approaches can also consider large perturbations. Such perturbations can occur in Earth system components such as the large ocean circulations or the Amazon rainforest~\citep{menck2013basin,dijkstra2005nonlinear}. The basin stability is an established algorithm focusing on the stability landscape of the entire phase space, while other nonlinear stability measures such as survivability~\citep{hellmann2016survivability}, stability threshold~\citep{klinshov2015stability}, constrained basin stability~\citep{van2016constrained} and topology of sustainable management~\citep{heitzig2016topology} approaches focus on the stability of parts of the state space or desired regimes in it. Therefore, basin stability computations are a first step that aims to quantify the stability of different attracting states, but do not aim to study potential desired regimes as would be required for the other mentioned methods. The concept of basin stability has been applied to many multistable systems. Examples comprise the Amazon rainforest~\citep{menck2013basin}, the stability in networks of power grids~\citep{schultz2014detours}, neuronal models~\citep{leng2016basin} and further nonlinear systems such as in coupled network systems~\citep{mitra2017multiple}, oscillators~\citep{rakshit2017basin,majhi2019emergence} or chimera states~\citep{rakshit2017basinchimera}. While basin stability can widely be applied, it has its limitations, for instance in cases where basins become too peculiar, e.g. for riddled basins with holes~\citep{schultz2017potentials}. Since this is not the case in our model of interacting climate tipping elements, we utilize basin stability in this work}.

An attractor $\mathcal{A}$ is defined as the minimal compact invariant set $\mathcal{A} \subseteq X$, where $X$ is the entire state space~\citep{milnor1985concept}. $B(\mathcal{A}) \subseteq X$ is the basin of attraction of $\mathcal{A}$ which comprises all states from which the system converges to $\mathcal{A}$. The basin stability or the basin volume $V(\mathcal{A})$ is then quantified as the probability that a system will return to a certain attractor $\mathcal{A}$ after a perturbation
\begin{equation}
    V_B(\mathcal{A}) := \int_R \mathbf{1}_{B(\mathcal{A})} d\mu \in [0, 1],
\label{eq:basin_stability}
\end{equation}
where $\mathbf{1}_{B(\mathcal{A})}$ is 1 in case $x \in B(\mathcal{A})$ and 0 otherwise. $\mu$ is a measure on the state space $X$ that encodes the relevance of a certain perturbation and our knowledge about the system. The estimation of the integral in Eq.~\ref{eq:basin_stability} can be difficult, but in our system it can be assumed that the estimation of the basin volume can be estimated via a Monte Carlo ensemble. The total volume of a basin of attraction is then measured as the fraction of simulations with randomly chosen initial conditions that end up in that certain attractor $N(\mathcal{A})$ over the total number of initial conditions $N(\Omega)$
\begin{equation}
    V(\mathcal{A}) = P(\mathcal{A}) = N(\mathcal{A})/N(\Omega).
\end{equation}
Here, $P(\mathcal{A})$ is the probability that a random initial condition ends up in the basin of attractor $\mathcal{A}$. To assign the basin volume $V(\mathcal{A})$ with the probability $P(\mathcal{A})$, it is required that the space of initial conditions is covered well and uniformly. Therefore in this work, it is necessary to extend the classical concept of basin stability since it is not only required to sample the space of initial conditions sufficiently well, but also to sample over the uncertainties in the model parameters themselves (see Tabs.~\ref{tab:appendix_1} and~\ref{tab:appendix_2}). Thus, we need to set up a very large-scale Monte Carlo ensemble of several billion ensemble members whose construction details can be directly found below.

\subsection{Monte Carlo ensemble to compute basin stability}
\label{met:monte}
\textcolor{black}{In order to apply the concept of basin stability in a meaningful way, the state space must be covered well enough. However, in this application, the parameters of the models have uncertainties themselves in the critical temperature thresholds and the interaction strength and structure. This means, we need a way of covering the many uncertainties in these various parameters as well as the state space itself. Therefore, it is necessary and useful to combine basin stability with a large scale Monte Carlo sample that covers an adequate extent of the phase space and parameter space. This is what explain in the hereafter.}\\

The basic Monte Carlo ensemble without the extension for basin stability is set up as follows: \textcolor{black}{for each pair of global mean temperatures (GMT) and interaction strengths d, there is a sample of size 100 constructed with initial conditions from the uncertainty range in $T_\text{limit,\ i}$ and $s_{ij}$ using a latin hypercube algorithm~\citep{pyDOE2013latin} (see Tabs.~\ref{tab:appendix_1} and~\ref{tab:appendix_2})}. Latin hypercube sampling is an extension to the usual random sampling and is used to improve the space coverage of initial conditions. Therefore, the space of initial conditions is separated into its dimensions, i.e., the number of different initial parameters (here 17, see Tabs.~\ref{tab:appendix_1} and~\ref{tab:appendix_2}). Then, it is secured that only one sample occurs in each axis hyperplane (compare to the N-rooks problem in mathematics). We apply this sampling procedure for each of the 27 different network setups that arise from the permutation (positive, negative, zero) of the three uncertain links (Amazon rainforest$\rightarrow$ENSO, AMOC$\rightarrow$Amazon and West Antarctic Ice Sheet$\rightarrow$AMOC, see Fig. 1A). This then leads to 2700 samples. These 2700 samples are computed for each global mean temperature increase up to 8~$^\circ$C above pre-industrial which can be reached in business as usual scenarios RCP8.5 extended from 2100 to 2500~\citep{schellnhuber2016right} in steps of 0.1~$^\circ$C and coupling constant d between 0.0 and 1.0 in steps of 0.02 accounting for 864.000 simulation runs.\\ 

\textcolor{black}{The extension of the Monte Carlo ensemble, integrating basin stability is detailed below:} the basin stability of the system for each of these 864.000 samples is computed by permuting the initial state of each of the five tipping elements within its limit, i.e., between the untipped (x=$-1.0$) and the tipped state (x=$+1.0$). The state variables of the five tipping elements result in a five dimensional state vector

\begin{align*}
    \text{vec}_\text{basin\ stability,\ i} = \{ &x_\text{Greenland\ Ice\ Sheet}(0),\ x_\text{AMOC}(0),\ x_\text{West\ Antarctic\ Ice\ Sheet}(0),\ \\ &x_\text{ENSO}(0),\ x_\text{Amazon\ rainforest}(0)\},
\end{align*}

where $\text{vec}_\text{basin\ stability,\ i} \in \left[ -1.0; +1.0 \right]$ for each tipping element. However, $\text{vec}_\text{basin\ stability,\ i}$ cannot be permuted in a completely random way, but each of its five dimensions needs to be permuted in an independent way since there is a strong nonlinearity at state equal zero for each of the five dimensions. \textcolor{black}{Of course, in principle if there would be infinite computational resources, we would not need to take this nonlinearity into account, but would be able to increase the size of the Monte Carlo ensemble even further. But since this is not the case, we need to ``manually'' account for this important nonlinear property.} This means that the sign of each state must be equally probable, i.e.:

\begin{align}
\begin{split}
P_\text{sign} &= P (-,-,-,-,-) = P(+,-,-,-,-) = P(-,+,-,-,-) = \\ &= … = P(+,+,+,-,+) = P(+,+,+,+,-) = P(+,+,+,+,+) = \\ &= 1/32 = 0.03125.
\end{split}
\label{eq:app:one}
\end{align}

This can be achieved when random starting conditions are drawn from each of the 32 combinations of $P_\text{sign}$. Hence, for each of the 32 combinations, we chose 10 different initial conditions ending up with 320 different settings. For the 320 randomly chosen perturbations (i.e., the initial conditions of the tipping elements), we again used a latin hypercube algorithm~\citep{pyDOE2013latin}. That means it fulfils the condition that each of the 32 different possible signs of the initial conditions in their five-dimensional subspace (one dimension for each tipping element) is covered equally often. \\

\begin{figure}[htbp]
\centering
\includegraphics[width=0.8\textwidth]{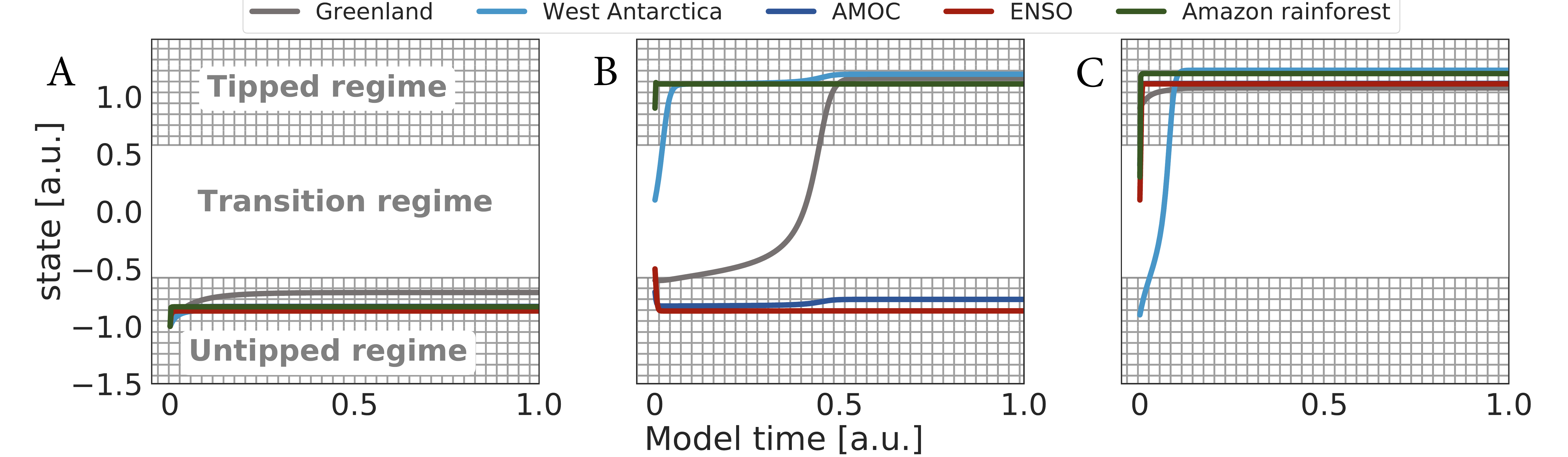}
\caption{A-C) Timelines at GMT=2.5~$^\circ$C above pre-industrial levels and $d=0.15$ with different initial conditions (ICs) as used to probe basin stability. A) $\text{IC} = [-1, -1, -1, -1, -1]$ (today's initial conditions, no element is tipped at $t=0$), B) $\text{IC} = [-0.6, -0.7, 0.1, -0.5, 0.9]$, C) $\text{IC} = [0.9, 0.4, -0.9, 0.1, 0.3]$.}
\label{fig:two}
\end{figure}

Altogether, we employ a very large ensemble of simulations to compute the basin stability of $N_\text{total} = 320 \cdot 864.000 = 3.569.184.000 \approx 3.6\cdot10^9$ samples. How the final state can depend on the initial conditions is shown exemplary for three timelines in Fig.~\ref{fig:two}A-C.

\subsection{Monte Carlo Basin Bifurcation Analysis}
\label{met:mcbb}
Coupling nonlinear ODEs as in the model described here, invokes the possibility of further types of bifurcations besides fold bifurcations. Here, we utilize Monte Carlo Basin Bifurcation Analysis~\citep{gelbrecht2020monte} to uncover system attractors and estimate their basins of attraction finding Hopf-Bifurcations and thus oscillating solutions converging to limit cycle attractors. MCBB is a novel, numerical approach to analyze multistable systems, quantify and track their asymptotic states in terms of their basins of attraction by utilizing random sampling and clustering methods. \textcolor{black}{Since MCBB is based on Monte Carlo ensembles and we are interested in a quantitative measure of interesting dynamical properties (here occurring limit cycles, i.e., Hopf-bifurcations), it is a well suited method for our purposes. It has also been applied to other nonlinear systems such as the Dodds-Watts model, the Kuramoto model or Stuart-Landau oscillators~\citep{gelbrecht2020monte}.}

MCBB aims to find classes of attractors that collectively share the largest basins of attractions of the system. Similar attractors, at different parameter values, have to share similar values of invariant measures $\rho$ and the difference of theses measures has to smoothly vanish if the parameter difference goes to zero. If this is the case they are regarded as being part of the same class of attractors. $N_{tr}$ trajectories of the system, here 140 000, with randomized initial conditions and parameters are integrated. In order to identify the different classes of attractors, suitable statistics $\mathbf{S}_{i}$ are measured on every system dimension for every trajectory, here, the mean, variance and the Kullbach-Leibler divergence to a normal distribution. Hence, for every statistic $i$, $\mathbf{S}_i$ is a $N_{tr}\times 5$ matrix. A distance matrix of each trajectory to each other is computed from these statics with 

\begin{align}
     D_{ij} = \sum_k^{3} w_k \sum_{l}^{5} |\mathbf{S}_{k,il} - \mathbf{S}_{k,jl}| + w_{4} | p^{(i)} - p^{(j)}|\label{eq:dist-direct}
\end{align}

where $p^{(i)}$ is the control parameter used to generate the $i$-th trajectory and $w_i$ are free parameters of the method, here $[1;0.5;0.5;1]$ which is the default recommendation for these parameters. This distance matrix is used as an input for a density-based clustering algorithm such as DBSCAN which can find if this notion of continuity between different trajectories exists and thus each cluster corresponds to a different class of attractors. For further details on MCBB, refer to \citep{gelbrecht2020monte}. 
When applying this to the conceptual model for climate tipping points, not only the different possible states of tipped elements are found, but also different classes of oscillating states induced by Hopf-Bifurcations are found. For the MCBB analysis, the parameter uncertainties were varied randomly within the same bounds as for the previously described basin computations. The initial conditions of five tipping elements were chosen to all start at $-1$, i.e., not tipped for the results presented in the main text, and at random between $-1$ and $+1$ for the results presented in the appendix. The computations are performed with the Julia library MCBB.jl.

\section{Results}
\label{results}
\subsection{Basin stability}
\label{res:basin}
\begin{figure}[htbp]
\centering
\includegraphics[width=0.8\textwidth]{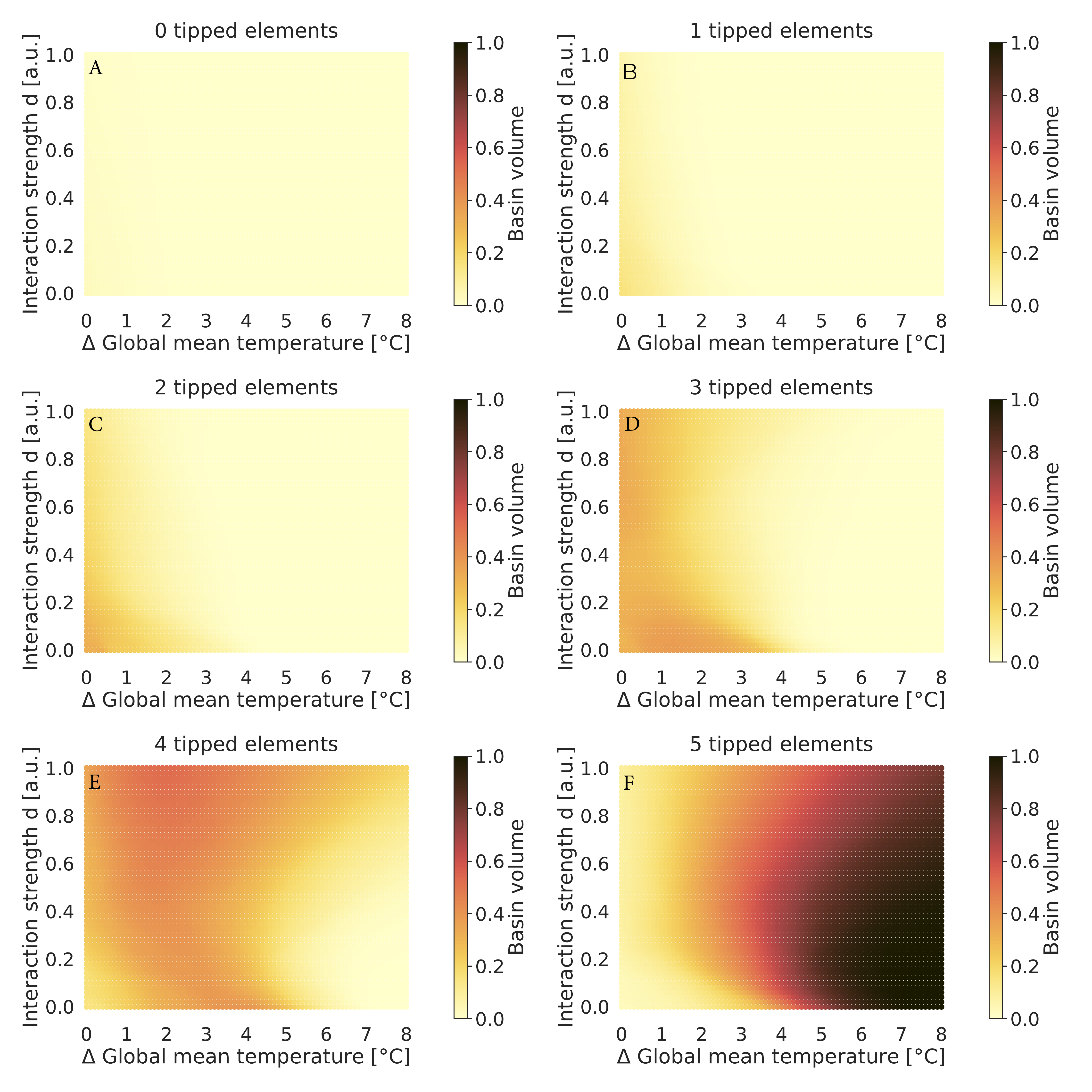}
\caption{Basin volume for each number of tipped elements in dependence of interaction strength and global mean temperature increase. A-F) Basin volume of 0 to 5 tipped elements.}
\label{fig:three}
\end{figure}

\textcolor{black}{We compute the basin stability of each potential state that could be governed by the network of five tipping elements.}. The present day state could be considered as some kind of safe state for the Earth system when all five tipping elements are in a negative state. On the other hand there could be a state where all five tipping elements reside in the positive, tipped state. In between there are intermediate scenarios, where some tipping elements already crossed their thresholds and others did not. In Fig.~\ref{fig:three}, we show the average basin stability for each of these six possible situations, i.e., with zero, one, two, three, four and five tipped elements. In this experiment, we perturbed the initial conditions of all tipping elements at the same time. The fraction of initial conditions that end up in the respective basin are plotted as the color. %\textcolor{black}{Importantly in this section, we only show the basin volume surplus, additional to the states that already start in that specific basin. Therefore, we subtract 1/32 for each computed point in the basin volume plots.}

In general, we observe that the size of the basin of attraction for higher global warming levels becomes larger for a higher number of tipped elements as would be expected. For high levels of warming, the basin of five tipped elements dominates. 

\textcolor{black}{For increasing interaction strength, the volume of the basins with three or less tipped elements decreases (Fig.~\ref{fig:three}A-D). Contrasting this, the basin volume with four tipped elements increases with increasing interaction strength, while the basin for five tipped elements first increases and then decreases again (Fig.~\ref{fig:three}E,~F).} the last issue is due to the strong negative feedback loop between the Greenland Ice Sheet and the AMOC. In such cases of high coupling, the AMOC tips, but safeguards the Greenland Ice Sheet which reaches the untipped regime for global mean temperature increases above 4~$^\circ$C and interaction strengths above 0.5. This poses a hypothetical scenario which would only be realistic if the interaction strength between Greenland and AMOC is very high, but this behavior has also been observed in experiments of tipping cascades earlier~\citep{wunderling2020interacting}.

For instance, in the basin volume plot of zero, one or two tipped elements, the number of states that equilibrate in this state is very small (Fig.~\ref{fig:three}A,~B and~C). For temperature increases above 2~$^\circ$C the associated basin volume is close to zero for all interaction strengths. At the same time, the size of the basin decreases for higher coupling strengths. 

The uncertainties of the basin volumes are quantified as standard deviation in the appendix (Fig.~\ref{fig:app:one}). We find that uncertainties generally increase for a higher amount of tipped elements as well as for higher interaction strengths. The standard deviation is highest for small temperature increases and high coupling strengths since here, the attractors depend on the initial conditions in terms of the critical temperature thresholds and initial coupling constants (see Tab.~\ref{tab:appendix_1}). The basin of four and five tipped elements show a regime of increased standard deviation for temperatures around 2-5~$^\circ$C above pre-industrial and interaction strength parameters of more than 0.2. This is probably due to the fact that in this regime the state of the Greenland Ice Sheet has a large variation because of its strong negative feedback loop to AMOC. Thus, whether this element tips, also depends a lot on the explicit initial conditions of the state as well as on parameters (see Tabs.~\ref{tab:appendix_1} and~\ref{tab:appendix_2}). Outside and around this regime, the uncertainty is smaller since either Greenland is not tipped with high certainty for lower temperature increases (below 2~$^\circ$C) or tipped with high certainty at higher temperature increases (above 5~$^\circ$C).

There exists a narrow range of global mean temperature increases when single tipping elements can transgress their state without triggering a tipping cascade. This range is mostly located below 1~$^\circ$C above pre-industrial for low coupling strength and well below 1~$^\circ$C for higher interaction strengths (see Fig.~\ref{fig:three}B and Fig.~\ref{fig:app:two}). If we separate this response into the respective singular tipping elements, we can see that above an interaction strength of 0.2-0.4, the Greenland Ice Sheet and ENSO cannot tip without causing a cascade due to their strong interactions links to AMOC or the Amazon rainforest, respectively (for more details see Appendix~\ref{app:three}).\\

Additionally, we investigate some important intermediate states in more detail, where some elements are in the tipped regime, while others are not. It was found that several tipping cascades of size two and three are more frequent than others, for instance a tipping cascade between the Greenland and the West Antarctic Ice Sheet is more likely than, for instance, a cascade between the AMOC and the Amazon rainforest~\citep{wunderling2020interacting}. Thus, we investigate the basin volume that corresponds to such cascades.

\begin{figure}[htbp]
\centering
\includegraphics[width=0.55\textwidth]{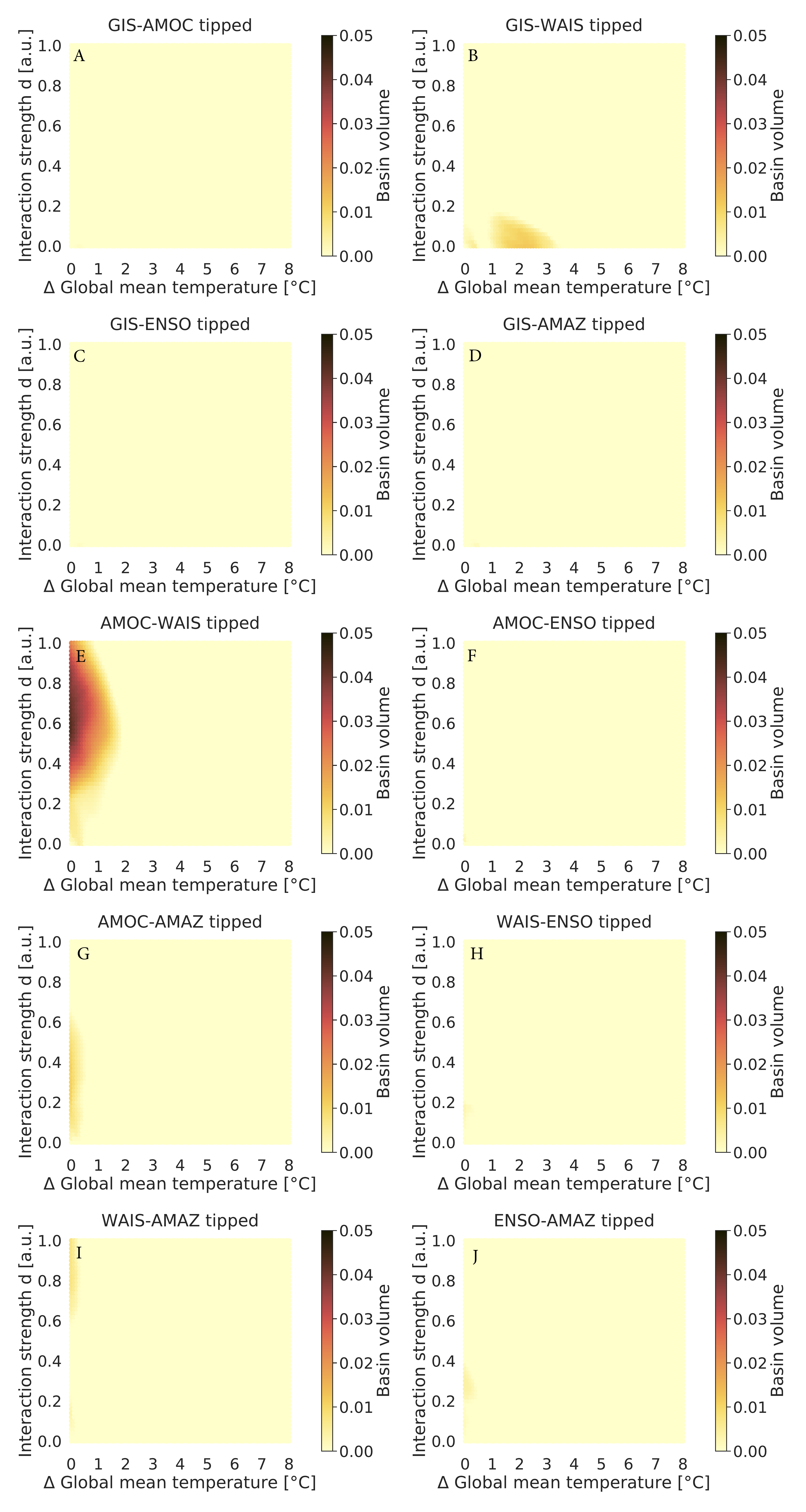}
\caption{Basin stability for exactly two tipped elements for all pairs of tipping elements (panel A to J). Most of the basin volumes are very small, but the joint basin of the Greenland and the West Antarctic Ice Sheet for low interaction strength and the joint basin of the AMOC and the West Antarctic Ice Sheet for high interaction strength and low temperature is increased. The abbreviations in the title stand for: Greenland Ice Sheet (GIS), West Antarctic Ice Sheet (WAIS), Atlantic Meridional Overturning Circulation (AMOC), El-Ni{\~n}o Southern Oscillation (ENSO) and Amazon rainforest (AMAZ).}
\label{fig:four}
\end{figure}

\begin{figure}[htbp]
\centering
\includegraphics[width=0.55\textwidth]{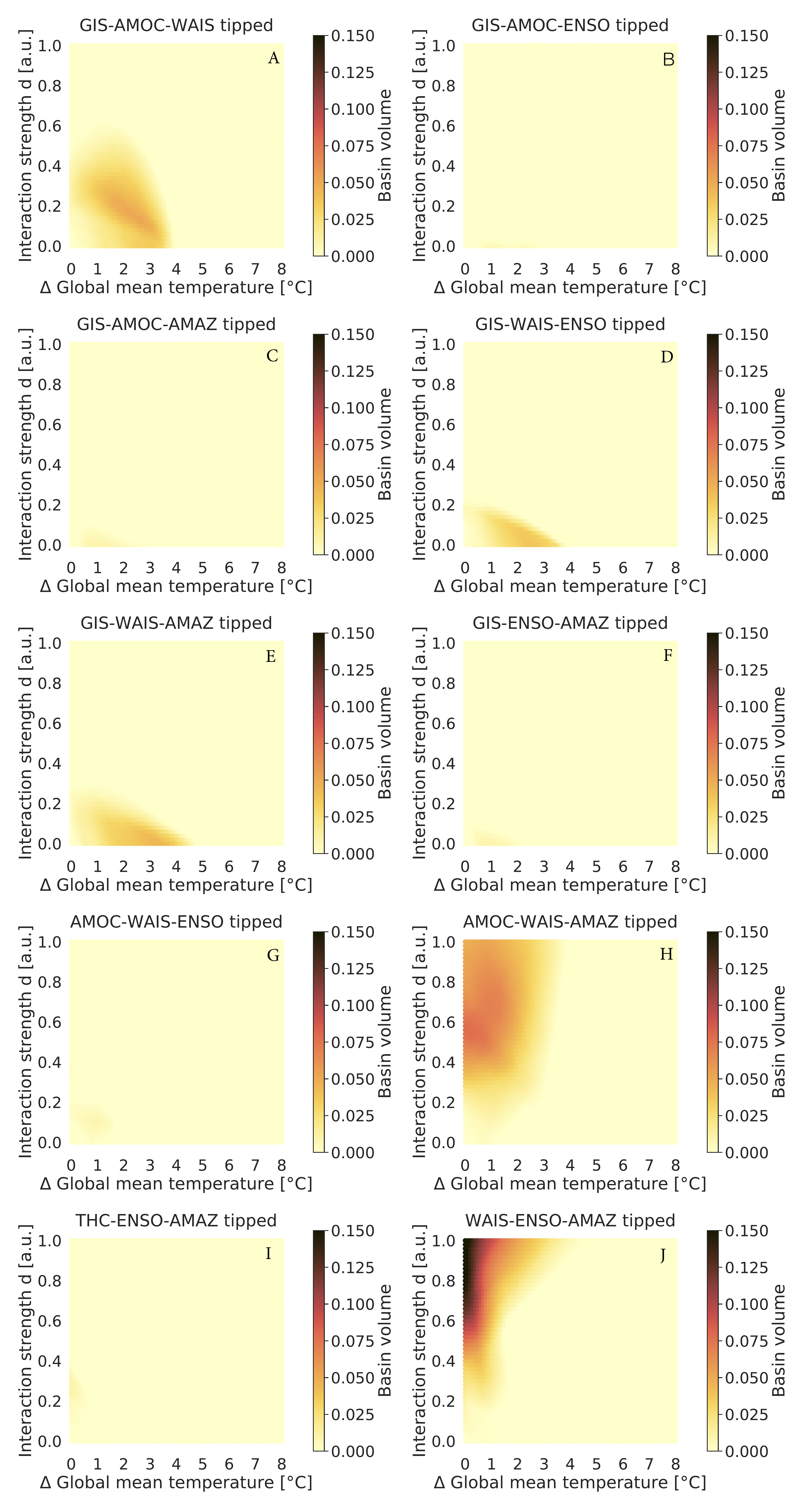}
\caption{Basin stability for exactly three tipped elements (panel A to J). The volume of basins is large, where the Greenland and the West Antarctic Ice Sheet are included, for low interaction strength and global warming levels of up to 4~$^\circ$C above pre-industrial.}
\label{fig:five}
\end{figure}

We find that the ice sheets appear to be of particular importance for the stability of the Earth system in our model because they have a high basin stability in both, when exactly two and exactly three elements are tipped (Figs.~\ref{fig:four} and~\ref{fig:five}). Although a potential disintegration of the ice sheets can take several centuries up to millennia, states including tipped ice sheets seem to be more stable than states without tipped ice sheets. This is also consistent with the earlier result that the large ice sheets are the initiators of many cascades in the studied model~\citep{wunderling2020interacting}.

In case exactly two elements are tipped (Fig.~\ref{fig:four}), the basin of the Greenland and the West Antarctic Ice Sheet is the only one which has increased basin volume for low interaction strength and global warming levels of 1-3~$^\circ$C above pre-industrial. This would represent a scenario in which, both, the Greenland Ice Sheet as well as the West Antarctic Ice Sheet are triggered and become ice free on long time scales without a tipping of the AMOC. This could for example be the case when global warming is higher than necessary to safeguard the large ice sheets, but low enough such that the time of their disintegration is slow enough such that the freshwater input into the AMOC does not stop their functioning.

In parallel, if exactly three elements are tipped, the combinations that include the Greenland and the West Antarctic Ice Sheet have a higher basin stability at low interaction strength. Here, global warming levels are up to 4~$^\circ$C above pre-industrial (Fig.~\ref{fig:five}).

\begin{figure}[htbp]
\centering
\includegraphics[width=\textwidth]{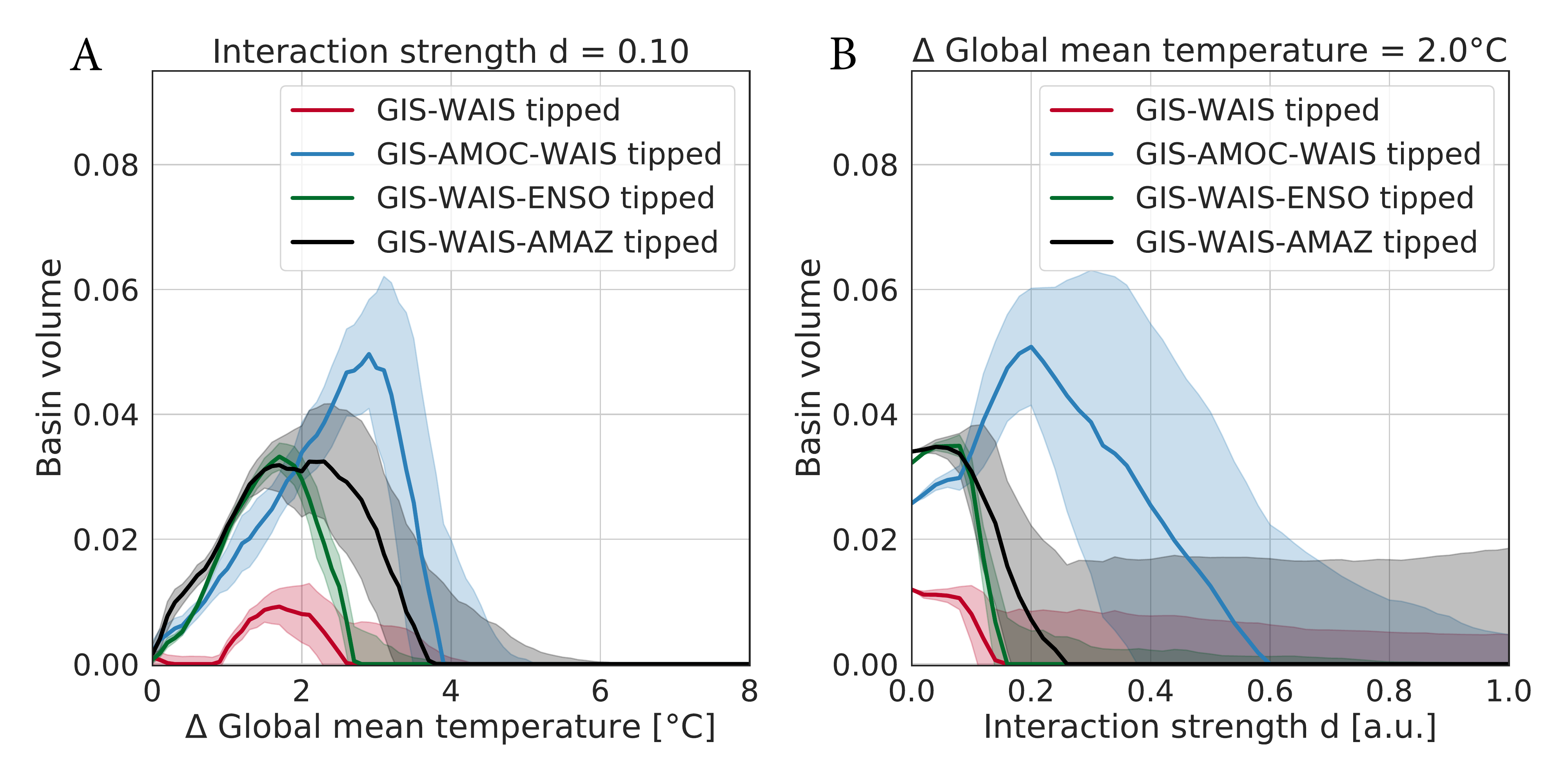}
\caption{\textcolor{black}{Basin volume for selected tipping elements residing in the tipped and untipped regime for A) an interaction strength of $d=0.10$ and B) a global mean temperature increase of 2.0~$^\circ$C above pre-industrial levels. The shaded colors represent the standard deviation arising from the 27 different network setups due to the structural uncertainties of the network of tipping elements (compare Sect.~\ref{met:monte} and see Fig.~\ref{fig:one}A). The standard deviation increases with increasing interaction strength $d$ in panel B) since the equilibrium state depends on the initial conditions of the respective ensemble member. This equilibrium state is influenced significantly more for higher interaction strengths, but due to stabilizing, destabilizing and unclear links in the network of tipping elements, the change of the equilibrium states fluctuates more and shows a higher uncertainty. Therefore, the standard deviation increases. The same is valid for panel A) at a region where (parts) the critical temperature ranges $T_\text{limit,\ i}$ are located for many of the tipping elements, that is mainly between 2-4~$^\circ$C above pre-industrial.}}
\label{fig:six}
\end{figure}

\textcolor{black}{We compare the basin volumes of these scenarios, where exactly two or three tipping elements are in the tipped regime and the large ice sheets are among these tipped elements (see Fig.~\ref{fig:six}). We observe that the basin volume is highest between 1-4~$^\circ$C above pre-industrial levels for an interaction strength of 0.1. We find that the basin volume is largest at intermediate interaction strengths $d$ (mainly below 0.2) for a global mean temperature increase of 2~$^\circ$C above pre-industrial levels. We also reveal that the basin volume for two tipped ice sheets (red curve) is lower than for exactly three tipped elements including the two ice sheets (other curves). Since many basin volumes of exactly two or three tipped elements are very close to zero (see Figs.~\ref{fig:four} and~\ref{fig:five}) and the basin volumes including tipped ice sheets are different from zero, this emphasizes again that the ice sheets could be of special interest for the resilience of the Earth system with respect to tipping dynamics.}

Furthermore, some basin volumes are increased for low to intermediate levels of global warming and high interaction strengths (above $\approx0.5$). It is likely that such scenarios are less realistic since, either such a low increase of the global mean temperature is improbable, or such high interaction strength would pose the unlikely scenario that interactions are as important as the individual dynamics of the tipping elements. This would be the case when the interaction strength $d$ approaches $1.0$ (see Figs.~\ref{fig:four}, ~\ref{fig:five} and compare to Eq.~\ref{eq:one}).

\newpage

\subsection{Oscillatory states}
\label{res:osci}
\begin{figure}[htbp]
\centering
\includegraphics[width=0.8\textwidth]{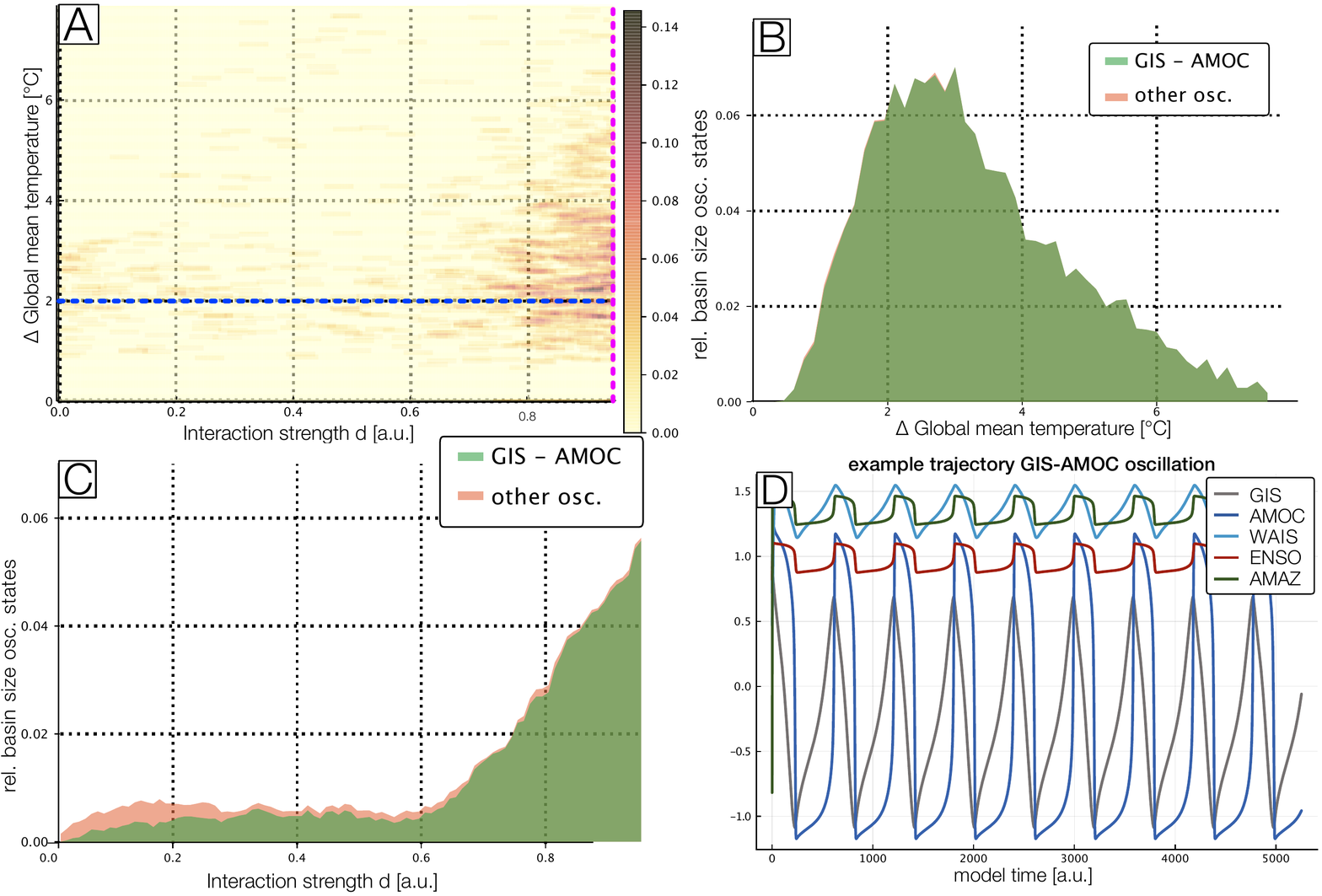}
\caption{Oscillating states in case the initial conditions of all tipping elements are $[-1, -1, -1, -1, -1]$. For random initial conditions, see Fig.~\ref{fig:app:three}. A) Occurrence of oscillating states with respect to global warming and interaction strength parameter d. B) Dependence of limit cycles and their main type on the temperature increase at high interaction strength (dashed magenta line in panel A). C) Dependence of limit cycles and their main type on the coupling strength at a temperature increase of 2~$^\circ$C above pre-industrial (dashed blue line in panel A). D) Example time series for a limit cycle of AMOC and GIS.}
\label{fig:seven}
\end{figure}

Furthermore, from the basin stability results we aim to separate off limit cycle attractors in the state space. The results from MCBB (Monte Carlo Basin Bifurcation~\citep{gelbrecht2020monte}) identify the parameter regimes where Hopf Bifurcations occur and the tipping elements start to show Kadyrov oscillations. Such Kadyrov oscillations have already been found in the early literature on dynamical systems of the CUSP type~\citep{abraham1991computational}. As shown in Fig.~\ref{fig:seven} for initial conditions at $-1$ for all tipping elements, this is most prominently the case for large interaction strengths and medium temperature increase values. Here, about every tenth solution is oscillating. This is due to the fact that uncertainties are largest in these regimes. For smaller interaction strength values, limit cycles can still occur but are much rarer with an occurrence at about 1\% of all solutions. Of all these limit cycle oscillations almost all (95\%) have a significant amplitude (Standard deviation $>$ 0.1) in at least one tipping element. The most common limit cycles are simultaneous oscillations of AMOC and GIS as shown in Fig.~\ref{fig:seven}D. They make up about 86\% of all oscillating states found. The reason for this predominant oscillation is that there is a strong negative feedback loop between the Greenland Ice Sheet and the AMOC via freshwater input from Greenland that weakens the AMOC, while on the other side a weaker AMOC cools the northern hemisphere~\citep[see e.g.][]{caesar2018observed,kriegler2009imprecise}. Still, whether such oscillations could indeed exist in the climate system remains speculative, but in principle there is evidence of oscillatory behavior in paleo data of the Earth system~\citep{ditlevsen2020crossover,crucifix2012oscillators}.

\newpage

\section{Discussion}
\label{disc}
\textcolor{black}{We find that the only dominating stable state in the long term, for large temperature increases around and above 4.0~$^\circ$C above pre-industrial levels, is the one with four or five tipped elements. Our results emphasize that the ice sheets could be of special importance for the stability of the climate system regarding their increased basin volume in case more than one element is tipped. Based on the known interactions from Kriegler et al.~\citep{kriegler2009imprecise} this makes sense, since the interactions between the ice sheets, especially from Greenland to West Antarctica, are strong due to potentially rising sea level that might cause grounding line retreat~\citep{sayag2013elastic}.} 

\textcolor{black}{Of course, the ice sheets interact with global modes of ocean variability like the AMOC and reduce its overturning strength, but in our model these interactions are not sufficient to tip the AMOC over in many cases. These states with disintegrated ice sheets are especially relevant exhibiting a high basin volume for intermediate climate warming scenarios consistent with the climate target of the Paris Agreement that aims at limiting global warming to well below 2~$^\circ$C above pre-industrial levels~\citep{paris2015}. Limit cycle oscillations between the tipping of some elements have been detected at some rare parameter configurations, mainly between the Greenland Ice Sheet and the AMOC. Although it remains unclear whether such (Kadyrov) oscillations have occurred in the climate system, they point to possible relevant internal modes of variability in the climate system. In principle such limit cycle behavior could have played a role in paleo climate dynamics such as in the Pleistocene ice age cycles~\citep{ditlevsen2020crossover,crucifix2012oscillators}. Further, the individual dynamics are not the sole determinant of the final state of the tipping elements since the network effects can cause additional tipping events. Through this network interaction, it is therefore possible that cascades of tipping events emerge, even before the actual critical temperature threshold for some of the tipping elements is reached~\citep{wunderling2020interacting}.}

\section{Conclusion}
\label{conc}

%SUMMARY PART
\textcolor{black}{In this work, we study a conceptual model of five climate tipping elements based on a system of coupled, nonlinear differential equations. We investigate the stability of different dynamical regimes with respect to its stable states applying the concept of basin stability using a very large-scale Monte Carlo simulation of more than 3.5 billion ensemble members. Following that approach, we are able to propagate the numerous uncertainties thoroughly which are associated with the critical temperature thresholds and interaction strengths. With a Monte Carlo basin bifurcation analysis tool, we detected oscillatory states within our system.}

\textcolor{black}{We observe that the largest basin volume is that of the basin, where all five tipping elements are in the transgressed state, especially for large levels of global warming. We also detect that the ice sheets are of special importance for the stability of states, where the large cryosphere components reside in the transgressed state, while the other tipping elements do not. We also detect Hopf-bifurcations for few parameter configurations (0.6\%), mainly taking place between the Greenland Ice Sheet and the AMOC (86\%).}\\

%LIMITATIONS PART
\textcolor{black}{Our complex dynamical networks approach strongly simplifies the nature of tipping elements as well as their interaction structure. However, it can serve to integrate simplified concepts of tipping elements until coupled, process-based models are developed that can resolve the respective nonlinearities in the Earth system in more detail since current state-of-the-art Earth system models cannot yet model all these nonlinearities due to a lack of comprehensive process-understanding and computational constraints. It is further important to note that some studies have hypothesized that major changes in ENSO are possible~\citep{timmermann2005enso,dekker2018cascading} based on conceptual models~\citep{timmermann2003nonlinear,zebiak1987model}, but however, whether this is evidence for a permanent and potentially even irreversibly tipped ENSO remains uncertain and debated. Surely, ENSO exerts strong feedbacks onto the climate system that will increase if major El-Ni\~{n}o events become more frequent, for instance through strong drying trends over Amazonia. Furthermore, in earlier research we found that the main results of our model remain robust under the omission of ENSO such that we decided to investigate the more complex case and included ENSO here, even though the use of Eq.~\ref{eq:one} is only a topologically equivalent dynamical equation~\citep[for more details see][]{wunderling2020interacting}. While some literature studies present ENSO among the list of potential tipping elements~\citep{lenton2008tipping,schellnhuber2016right,kriegler2009imprecise}, it still remains uncertain whether ENSO is a tipping element in a strict sense.}\\

%OUTLOOK PART
%Future research should try to identify possible feedback processes that might impact the stability of the tipping elements ad- or conversely due to the disintegration of some elements. This might reveal important \textit{intervention points}, where the Earth system might be governed in a more or less safe way. Hereby, emulator-like models for tipping elements could help to identify such stabilizing points, as they already exist for non tipping element related contexts (e.g. MAGICC-6 or Pyhector)~\citep{meinshausen2011emulating,meinshausen2011emulating_two,willner2017pyhector}. %\textcolor{red}{COMMENT: I am not sure about this very last outlook paragraph? How can we improve this part? Oder ganz weglassen?}

\textcolor{black}{Overall, our network approach can easily be adapted to further tipping elements as soon as their interaction structure would be understood. It is also possible to probe the effect of different structural interaction hypotheses to further tipping elements within the scope of an uncertainty analysis, as has already been performed here for three interaction links. Further, the results of our study motivate that it could be worthwhile to look into the dynamics in more detail using process-detailed Earth system models. Especially the role of the large sheets in the stability landscape and oscillations between climate system components could be of interest. Even though, there is some knowledge about the interaction structure present in literature (see Sect.~\ref{met:interactions}), a new expert elicitation might be worthwhile because the knowledge about the interactions between the tipping elements has surely widened since the original expert elicitation from Kriegler et al. (2009)~\citep{kriegler2009imprecise}.}

\newpage \clearpage
\bibliographystyle{iopart-num}
\bibliography{bibliography}
\newpage \clearpage

\textbf{Acknowledgements}\\
This work has been carried out within the framework of PIK's FutureLab on Earth Resilience in the Anthropocene. N.W., M.G. and R.W. acknowledge the financial support by the IRTG 1740/TRP 2015/50122-0 project funded by DFG and FAPESP. N.W. is grateful for a scholarship from the Studienstiftung des deutschen Volkes. J.F.D. is grateful for financial support by the Stordalen Foundation via the Planetary Boundary Research Network (PB.net), the Earth League’s EarthDoc program and the European Research Council Advanced Grant project ERA (Earth Resilience in the Anthropocene, ERC-2016-ADG-743080). We are thankful for financial support by the Leibniz Association (project DominoES). The authors gratefully acknowledge the European Regional Development Fund (ERDF), the German Federal Ministry of Education and Research and the Land Brandenburg for supporting this project by providing resources on the high performance computer system at the Potsdam Institute for Climate Impact Research.\\

\textbf{Author contributions}\\
%N.W., R.W. and J.F.D. designed the study. N.W. conducted the model simulation runs and prepared the figures for the part on basin stability. M.G. conducted the simulation runs and prepared the figures for the part on oscillatory states. All authors discussed the results of the paper and N.W. led the writing with input from all authors.\\
\textcolor{black}{CRediT (Contributor Roles Taxonomy) statement:
\textbf{N.W.:} Conceptualization, Formal analysis, Investigation (Basin stability, Monte Carlo), Visualization (Introduction, Methods, Basin stability), Writing - Original Draft, Writing - Review \& Editing. \textbf{M.G.:} Formal analysis, Investigation (Monte Carlo Basin Bifurcation), Visualization (Oscillatory states), Writing - Review \& Editing. \textbf{R.W.:} Conceptualization, Writing - Review \& Editing, Supervision, Funding acquisition. \textbf{J.K.:} Writing - Review \& Editing, Funding acquisition. \textbf{J.F.D.:} Conceptualization, Writing - Review \& Editing, Supervision, Funding acquisition.}\\

\textcolor{black}{\textbf{Code and data availability}\\
The data that support the findings of this study are available from the corresponding author upon reasonable request. The code for the Monte Carlo ensemble construction and the conceptual Earth system that support the findings of this study are freely (3-clause BSD license) available on github under the following \href{http://dx.doi.org/10.5281/zenodo.4153102}{doi: 10.5281/zenodo.4153102}. The algorithm on the Monte Carlo Basin Bifurcation (MCBB) is available directly from the GitHub repository \href{https://github.com/maximilian-gelbrecht/MCBB.jl/}{https://github.com/ma-ximilian-gelbrecht/MCBB.jl/}. For the use of MCBB, please also confer~\citep{gelbrecht2020monte}.}\\

\textbf{Note on color maps}\\
This paper makes use of the conceptually uniform colormaps developed by~\citep{crameri2018geodynamic}.\newpage

\section*{Appendix}
\appendix
\counterwithin{figure}{section}
\counterwithin{table}{section}

\section{Parameter uncertainties}
\label{app:one}
In the following tables (Tabs.~\ref{tab:appendix_1} and~\ref{tab:appendix_2}), we list the critical temperatures $T_\text{limit,\ i}$ for the respective tipping element and the interactions between them together with their uncertainties.
\begin{table}[htbp]
\centering
\resizebox{\linewidth}{!}{
\begin{tabular}{|l|c|c|}%
\toprule
\textbf{Interaction} & \textbf{Link strength range $s_{ij}$ (a.u.)} & \textbf{Process} \\ \toprule
Greenland $\rightarrow$ AMOC & $\left[+1; +10\right]$ & Freshwater inflow \\
AMOC $\rightarrow$ Greenland & $\left[-1; -10\right]$ & AMOC breakdown, Greenland cooling \\
Greenland $\rightarrow$ West Antarctica & $\left[+1; +10\right]$ & Grounding line retreat \\
ENSO $\rightarrow$ Amazon rainforest & $\left[+1; +10\right]$ & Drying over Amazonia \\\hline

ENSO $\rightarrow$ West Antarctica & $\left[+1; +5\right]$ & Warming of Ross and Amundsen seas\\
AMOC $\rightarrow$ Amazon rainforest & $\left[\pm 2; \pm 4\right]$ & Changes in hydrological cycle \\
West Antarctica $\rightarrow$ AMOC & $\left[\pm 1; \pm 3\right]$ & Increase in meridional salinity gradient ($-$), \\
 &  & Fast advection of freshwater anomaly \\
 &  & to North Atlantic ($+$) \\\hline
 
AMOC $\rightarrow$ ENSO & $\left[+1; +2\right]$ & Cooling of North-East tropical Pacific with thermo- \\
 &  & cline shoaling and weakening of annual cycle in EEP \\
West Antarctica $\rightarrow$ Greenland & $\left[+1; +2\right]$ & Grounding line retreat \\
ENSO $\rightarrow$ AMOC & $\left[-1; -2\right]$ & Enhanced water vapor transport to Pacific \\
AMOC $\rightarrow$ West Antarctica & $\left[+1; +1.5\right]$ & Heat accumulation in Southern Ocean \\
Amazon rainforest $\rightarrow$ ENSO & $\left[\pm 1; \pm 1.5 \right]$ & Changes in tropical moisture supply \\
\bottomrule
\end{tabular}
}
\caption{Each interaction in the network of Fig.~\ref{fig:one} has a specific link strength range and a specific physical process that is connected to the respective interaction. The link strength ranges are computed from literature values~\citep{lenton2013origin,kriegler2009imprecise} such that they can be used in Eq.~\ref{eq:one}. For a more in depth description please be referred to Wunderling et al. (2020)~\citep{wunderling2020interacting}.}
\label{tab:appendix_1}
\end{table}

\begin{table}[htbp]
\centering
\begin{tabular}{|l|c|c|}%
%\resizebox{\linewidth}{!}{
\toprule
\textbf{Tipping element} & \textbf{$\mathbf{ T_\text{limit,\ i}}$ [$\mathbf{^\circ}$C]} & \textbf{$\mathbf{\tau_i}$ [a.u.]} \\ \toprule
Greenland & 0.8 -- 3.2 & 4900 \\
West Antarctica & 0.8 -- 5.5 & 2400 \\
AMOC & 3.5 -- 6.0 & 300 \\
ENSO & 3.5 -- 7.0 & 300 \\
Amazon rainforest & 3.5 -- 4.5 & 50 \\
\bottomrule
\end{tabular}
%}
\caption{Critical temperature range $T_\text{limit,\ i}$ of the five tipping elements as taken from the literature~\citep{schellnhuber2016right}, see also Eq.~\ref{eq:one}. The typical tipping time scale $\tau_i$ is given in model years (in arbitrary units) since it is beyond the scope of this model to make predictions about the exact tipping times. However, certain differences in tipping times as used here can be decisive whether a tipping event occurs or not. For more information see Wunderling et al. (2020)~\citep{wunderling2020interacting}.}
\label{tab:appendix_2}
\end{table}

\section{More basin stability results}
\label{app:three}
\begin{figure}[htbp]
\centering
\includegraphics[width=0.8\textwidth]{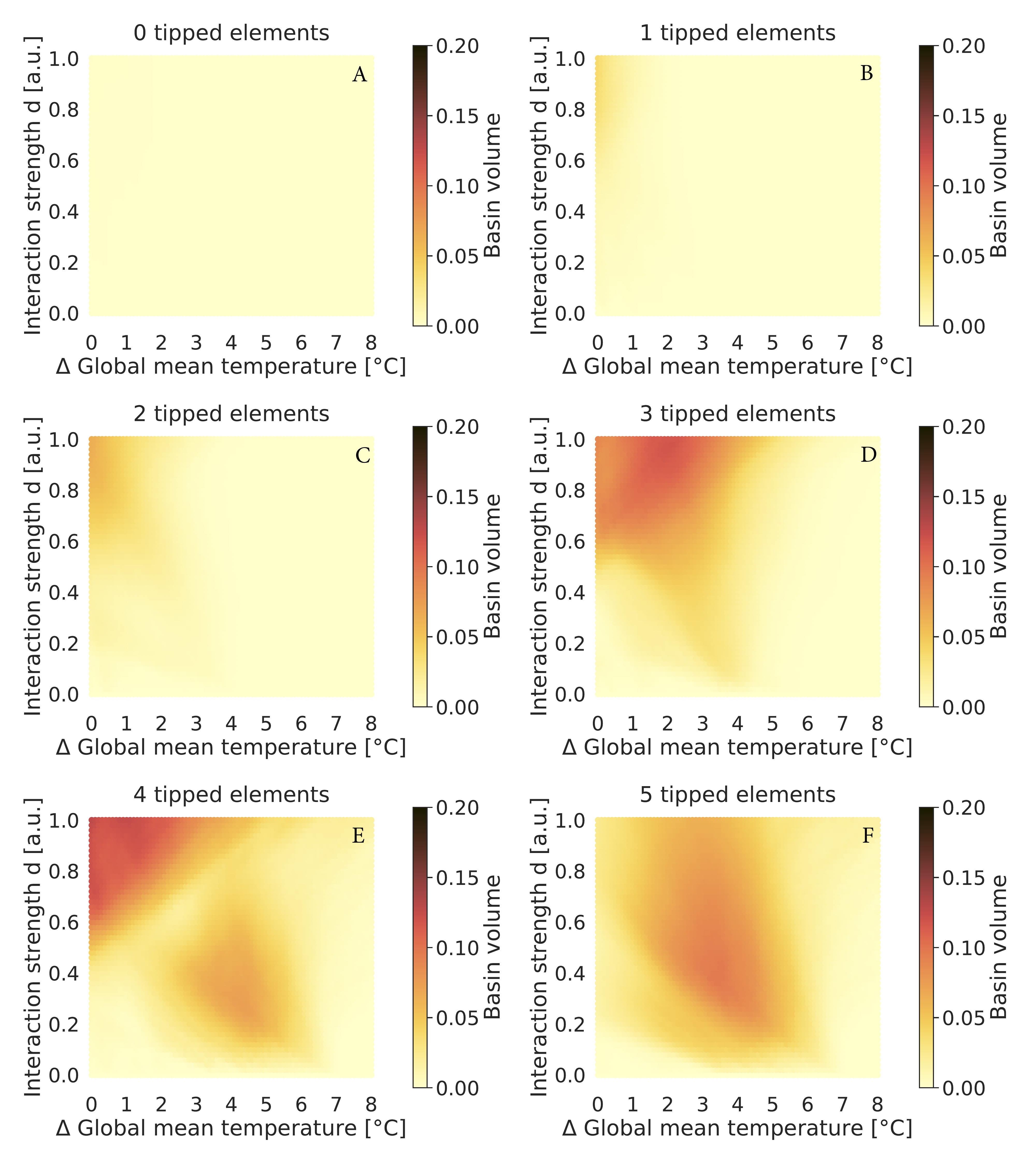}
\caption{Standard deviation over the 27 different network realizations of the basin volume normalized to one for each number of tipped elements (panels A-F) in dependence of interaction strength and global mean temperature increase. Mean values can be found in Fig.~\ref{fig:three}.}
\label{fig:app:one}
\end{figure}

\begin{figure}[htbp]
\centering
\includegraphics[width=0.8\textwidth]{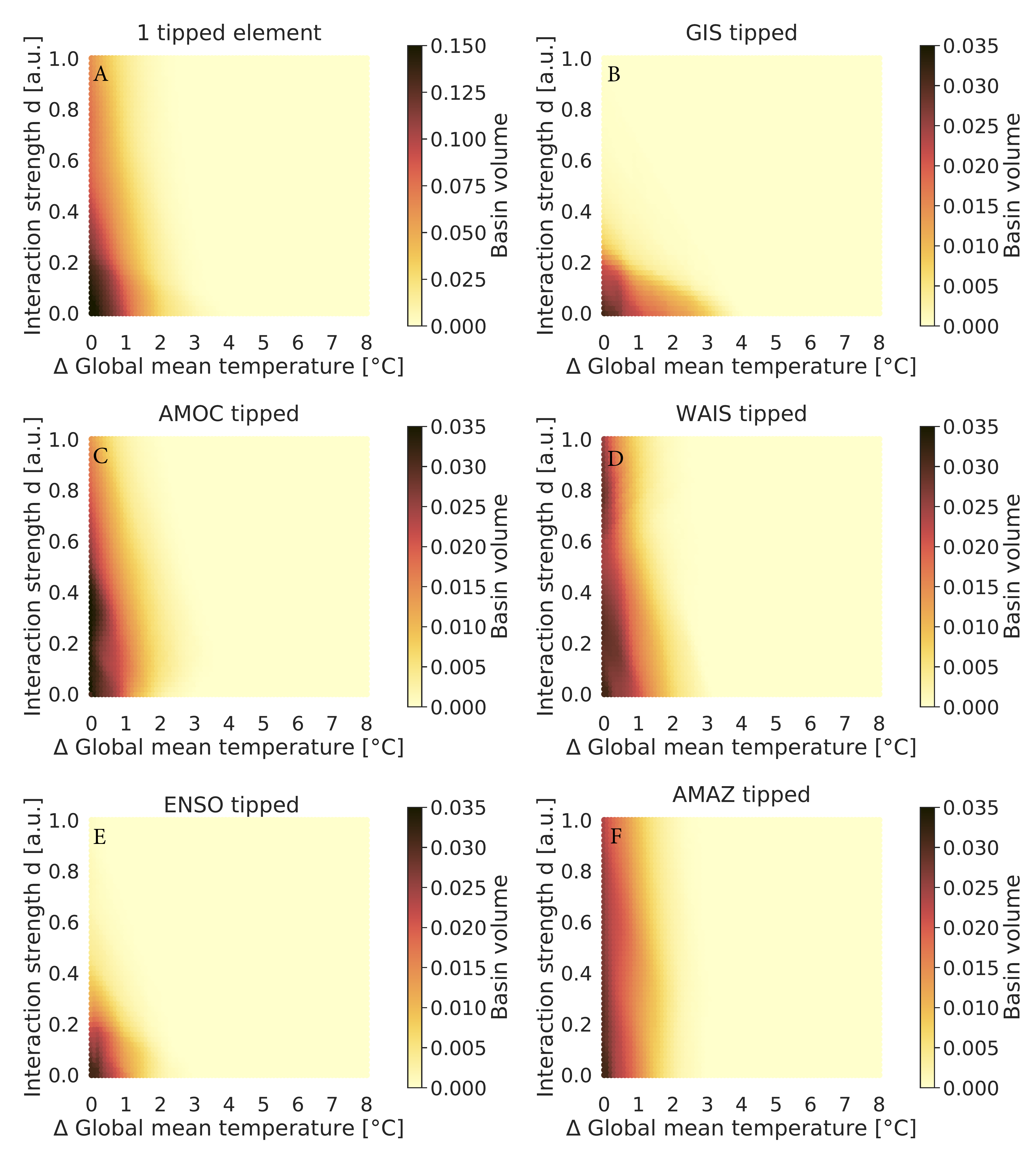}
\caption{Basin stability for single elements showing that the basin volume for the AMOC, WAIS (West Antarctic Ice Sheet) and AMAZ (Amazon rainforest) are qualitatively similar since it is possible that only this particular element tips. However, for ENSO and the GIS (Greenland Ice Sheet) this is not the case. It can onyl very rarely happen that these elements tip on themselves for high interaction strength even at low temperature increases since both of them possess a very strong link to another element that they would draw along into the tipped state. For ENSO, this is the Amazon rainforest and for the Greenland Ice Sheet it is the AMOC. Note that the color bar is different for panel A) to improve visibility.}
\label{fig:app:two}
\end{figure}

Here, we show the standard deviation of the basin volume for 0 to 5 tipped elements (Fig.~\ref{fig:app:one}) and the basin volume for one specific tipped element (Fig.~\ref{fig:app:two}) to complement the results from Fig.~\ref{fig:three}.

\section{Oscillatory regimes for random initial conditions}
\label{app:four}
\begin{figure}[htbp]
\centering
\includegraphics[width=\textwidth]{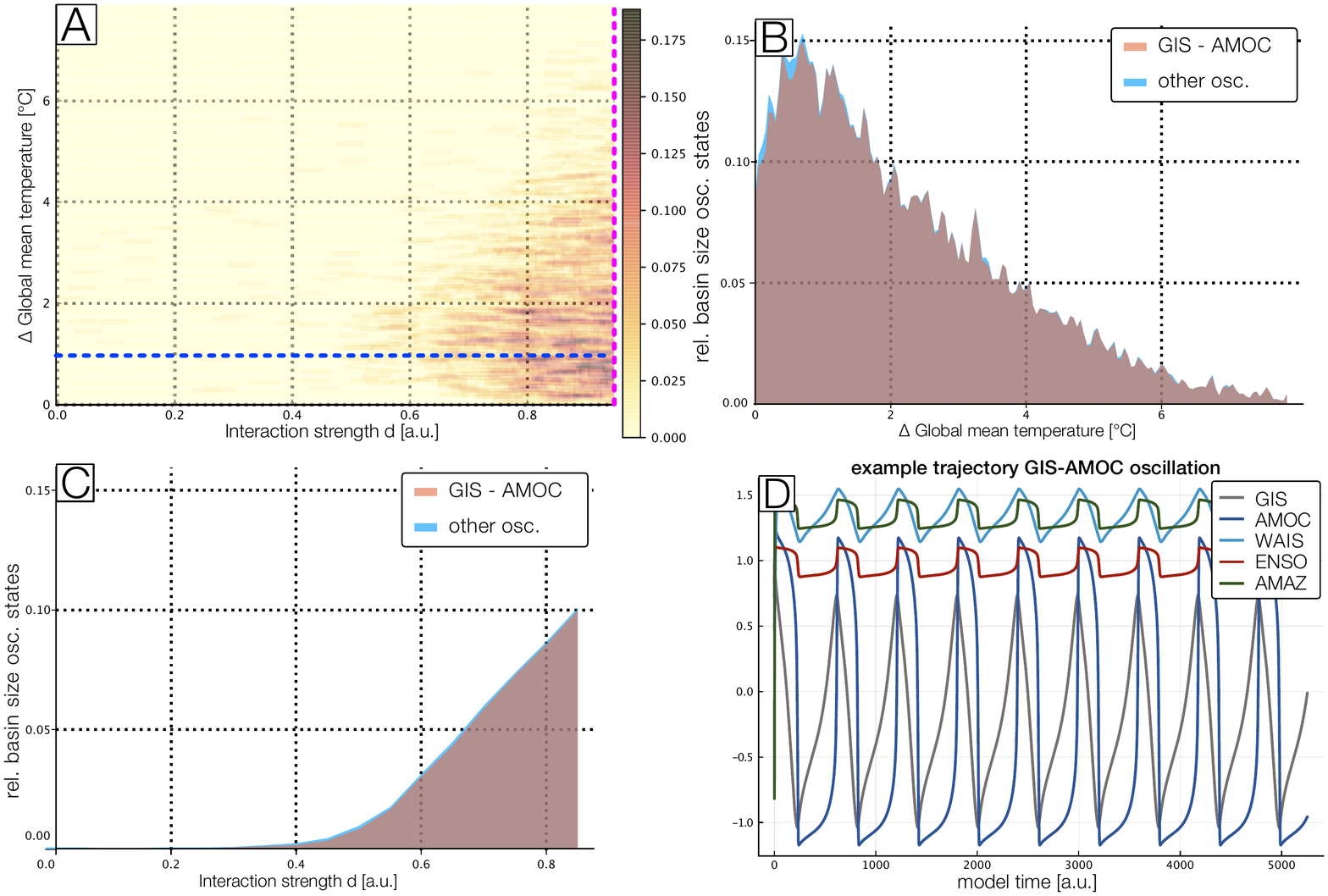}
\caption{Oscillating states for random initial conditions. Limit cycles occur more often than for initial conditions at -1 for all tipping elements. On the other side, the limit cycle oscillation between the AMOC and the Greenland Ice Sheet is still dominating.}
\label{fig:app:three}
\end{figure}

Here, we show the results of a Monte Carlo Basin Bifurcation analysis for random initial conditions (Fig.~\ref{fig:app:three}). We find that limit cycles occur more frequently when the initial conditions are randomly shuffled.

\end{document}